# Pushing limits of photovoltaics and photodetection using radial junction nanowire devices

Vidur Raj
*Mazumdar-Shaw Advanced Research Centre*
*University of Glasgow*
University Avenue, Glasgow G12 8QQ, UK
vidur.raj@glasgow.ac.uk

Yi Zhu
*Cambridge Graphene Centre*
*University of Cambridge*
Cambridge CB3 0FA United Kingdom
yz778@cam.ac.uk

Kaushal Vora
*Australian National Fabrication Facility*
*Research School of Physics, The Australian National University*
Canberra, ACT 2600, Australia
kaushal.vora@anu.edu.au

Lan Fu
*Australian Research Council Centre of Excellence for Transformative Meta-Optical Systems, Department of Electronic Materials Engineering, Research School of Physics, The Australian National University*
Canberra, ACT 2600, Australia lan.fu@anu.edu.au

Hark Hoe Tan
*Australian Research Council Centre of Excellence for Transformative Meta-Optical Systems, Department of Electronic Materials Engineering, Research School of Physics, The Australian National University*
Canberra, ACT 2600, Australia
hoe.tan@anu.edu.au

Chennupati Jagadish
*Australian Research Council Centre of Excellence for Transformative Meta-Optical Systems, Department of Electronic Materials Engineering, Research School of Physics, The Australian National University*
Canberra, ACT 2600, Australia
chennupati.jagadish@anu.edu.au

*Abstract*— Nanowire devices have long been proposed as an efficient alternative to their planar counterparts for different optoelectronic applications. Unfortunately, challenges related to the growth and characterization of doping and p-n junction formation in nanowire devices (along axial or radial axis) have significantly impeded their development. The problems are further amplified if a p-n junction has to be implemented radially. Therefore, even though radial junction devices are expected to be on par with their axial junction counterparts, there are minimal reports on high-performance radial junction nanowire optoelectronic devices. This paper summarizes our recent results on the simulation and fabrication of radial junction nanowire solar cells and photodetectors, which have shown unprecedented performance and clearly demonstrate the importance of radial junction for optoelectronic applications. Our simulation results show that the proposed radial junction device is both optically and electrically optimal for solar cell and photodetector applications, especially if the absorber quality is extremely low. The radial junction nanowire solar cells could achieve a 17.2% efficiency, whereas the unbiased radial junction photodetector could show sensitivity down to a single photon level using an absorber with a lifetime of less than 50 ps. In comparison, the axial junction planar device made using same substrate as absorber showed less than 1% solar cell efficiency and almost no photodetection at 0 V. This study is conclusive experimental proof of the superiority of radial junction nanowire devices over their thin film or axial junction counterparts, especially when absorber lifetime is extremely low. The proposed device holds huge promise for III-V based photovoltaics and photodetectors.

## I. Introduction

Since the emergence of nanotechnology, III-V nanowires have been proposed as an efficient, low-cost alternative to their thin film counterparts for different optoelectronic applications, including LEDs/lasers, photovoltaics, and photodetectors.[1] Nanowires have several advantages compared to their planar counterparts, such as high absorption and emission per unit volume, significantly higher growth rate, facile strain relaxation allowing for heteroepitaxy, and increased defect tolerance, which makes them interesting for optoelectronics.[2] Nanowires also allow for new device architectures which would otherwise be difficult to implement in planar devices or cannot be implemented at all.[3]

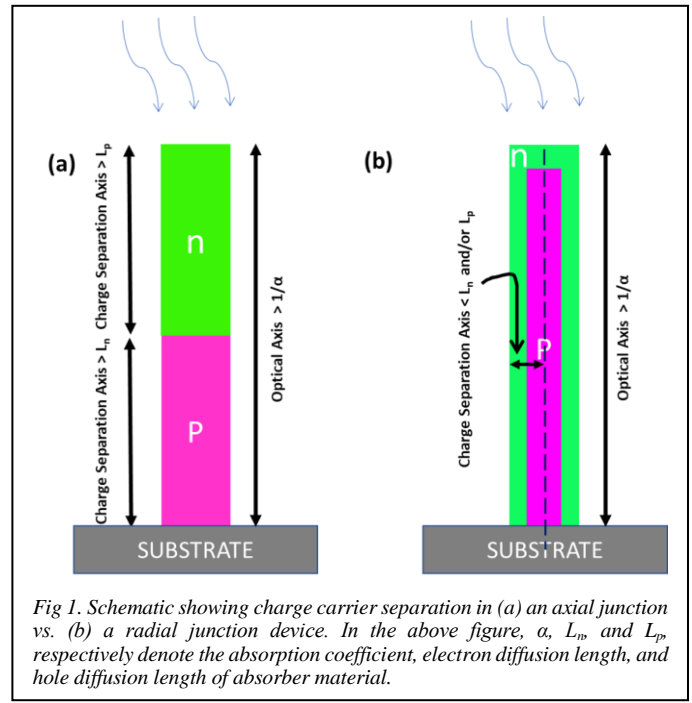

Fig 1. Schematic showing charge carrier separation in (a) an axial junction vs. (b) a radial junction device. In the above figure, α, $L_n$, and $L_p$, respectively denote the absorption coefficient, electron diffusion length, and hole diffusion length of absorber material.

Depending on the dominant axis of charge carrier separation, nanowire devices can be categorized into axial or radial junction devices.[4, 5] Almost all planar/nanowire devices are axial junction devices, and they have been widely studied.[2] First proposed in 2005, radial junction nanowire devices have been presented as an efficient alternative to their axial junction counterpart to circumvent the performance loss due to the small diffusion length of active material, as is mostly the case with nanowires.[6] In a radial junction nanowire device, the axis of charge carrier separation is decoupled from their optical axis, allowing for separate optimization of device electronic and optical behavior, as schematically illustrated in figure 1.

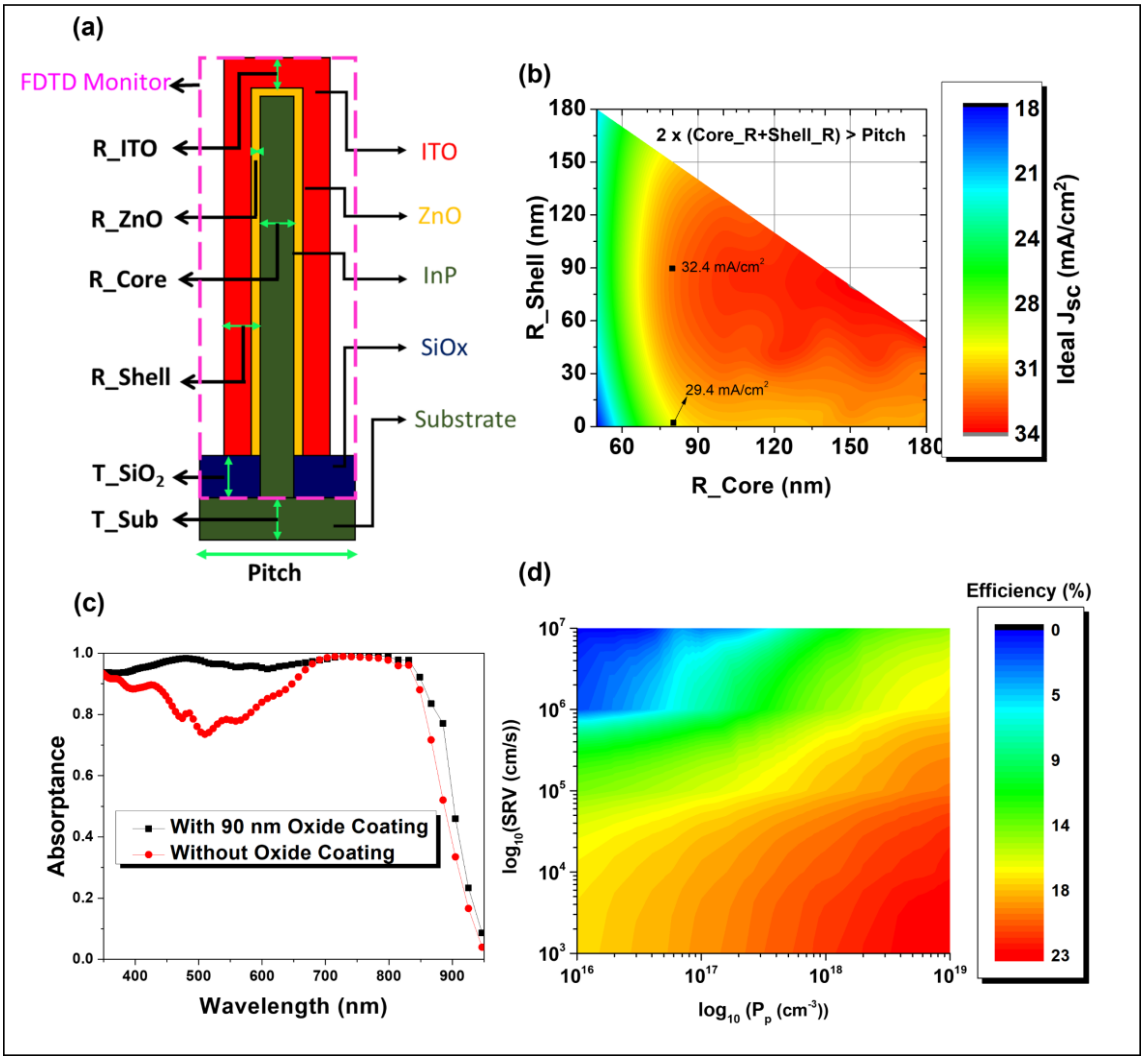

*Fig 2. (a) A 2-D schematic with proposed radial junction nanowire devices, with clearly labeled dimensions of device structures. (b) shows deposition of about 90 nm of oxide over bare InP nanowire can help achieve about 32.4 mA/cm$^2$ for filling ratio as low as 0.09. In comparison, bare InP could only achieve 29.4 mA/cm$^2$. (c) shows absorption in 160 nm wide (600 nm pitch) InP nanowire with and without 90 nm of the oxide coating. (d) shows that even when material lifetime is 50 ps, one can achieve more than 23% efficiency for doping higher than $10^{18}$ cm$^{-3}$ and a surface recombination velocity of about $10^4$ cm/s. [Figure 2 has been reproduced (adapted) with permission from ref [5]. Copyright © 2019 IEEE Publishers]*

Unfortunately, controlling the device behavior of bottom-up grown nanowire devices has been challenging because of the difficulties to achieve controlled doping and, by extension, to form a controlled p-n junction. Another big issue in the fabrication of high-performance nanowire devices has been their large surface area, which makes surface recombination a significant problem. Moreover, the growth of conventional passivation layers, which consist of wide band gap, heavily doped, epitaxial window layer, remains extremely challenging, in particular for core-shell nanowires, which have been a primary hindrance to realizing high-performance radial junction nanowire devices. Therefore, even though radial junction has been predicted as an efficient alternative to their axial or planar junction devices, experimentally, axial junction devices have shown superior performance. Since the emergence of nanotechnology, III-V nanowires have been proposed as an efficient, low-cost alternative to their thin film counterparts for different optoelectronic applications, including LEDs/lasers, photovoltaics, and photodetectors.[1]

Nanowires have several advantages compared to their planar counterparts, such as high absorption and emission per unit volume, significantly higher growth rate, facile strain relaxation allowing for heteroepitaxy, and increased defect tolerance, which makes them interesting for optoelectronics.[2] Nanowires also allow for new device architectures which would otherwise be difficult to implement in planar devices or cannot be implemented at all.[3]

Only recently, we showed both theoretically and experimentally the absolute superiority of radial junction nanowire devices over axial junction devices when the absorber lifetime (i.e., low diffusion length) is extremely low.[5, 7-9] Because of the reasons discussed above, we use top-down etched p-type InP nanowires to fabricate radial junction devices instead of the bottom-up approach. The p-n junction is formed by depositing a conformal coating of n-type aluminum zinc oxide (AZO)/ZnO on the p-type InP nanowires. ZnO oxide acts as an electron selective contact and also passivates InP nanowire by creating a thin interface layer of InPO$_x$,[4, 10] whereas AZO acts as a transparent conducting oxide. We showed that using this radial junction architecture, an absorber layer with less than 100 ps bulk

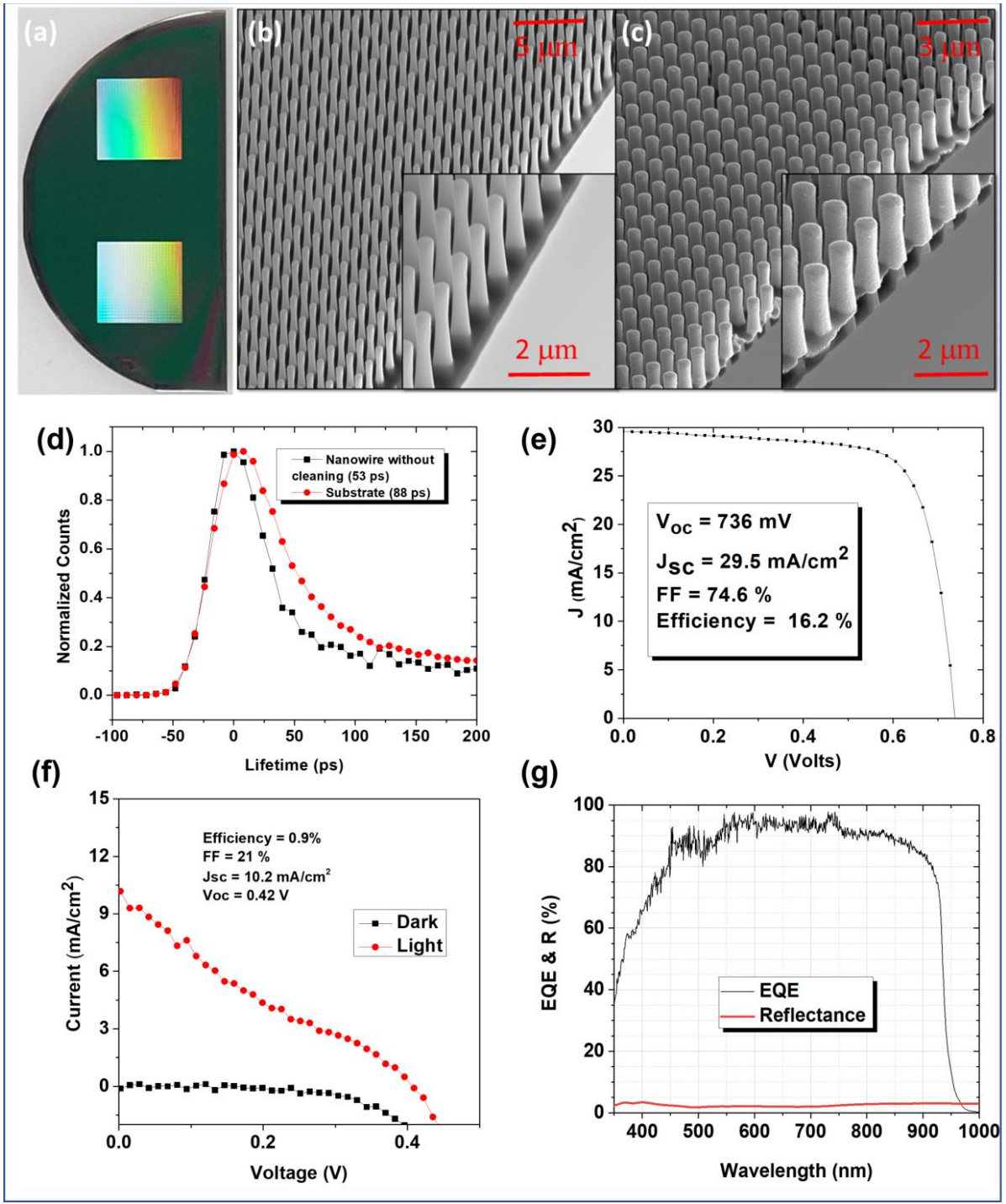

Fig 3. (a) An example of wafer scale e-beam lithography patterned nanowire arrays. (b) and (c) respectively shows the SEM image of top-down etched InP with and without AZO/ZnO shell. (d) TRPL measurement of a p-type substrate and top-down etched InP nanowires. (e) Light JV curve of radial junction nanowire solar cells. (f) Light and dark JV of a planar junction solar cell using a substrate with a lifetime of less than 100 ps. (g) EQE of the radial junction nanowire solar cell array. *[Figure 3 has been reproduced (adapted) with permission from ref [7]. Copyright © 2019 ACS publications]*

lifetime could be transformed to achieve high-performance solar cells with a world record short-circuit current of 31.3 mA/cm$^2$ and photodetectors which can achieve broadband single photon level detection at room temperature. In comparison, planar junction solar cells made of the same absorber layer had less than 1% efficiency and showed extremely low-efficiency photodetection. This report summarizes some of these results on the fabrication of radial junction nanowire devices for application in solar cells and single-photon detectors.

## II. NANOWIRE SOLAR CELL – SIMULATION AND FABRICATION

### A. Simulations

Here, we only summarize the results on the simulation and fabrication of radial junction nanowire solar cells. For a more detailed discussion, readers may refer to references [5, 7-9]. Figure 2(a) shows a 2-D schematic of radial junction nanowire solar cell. It consists of heavily p-type doped, top-down

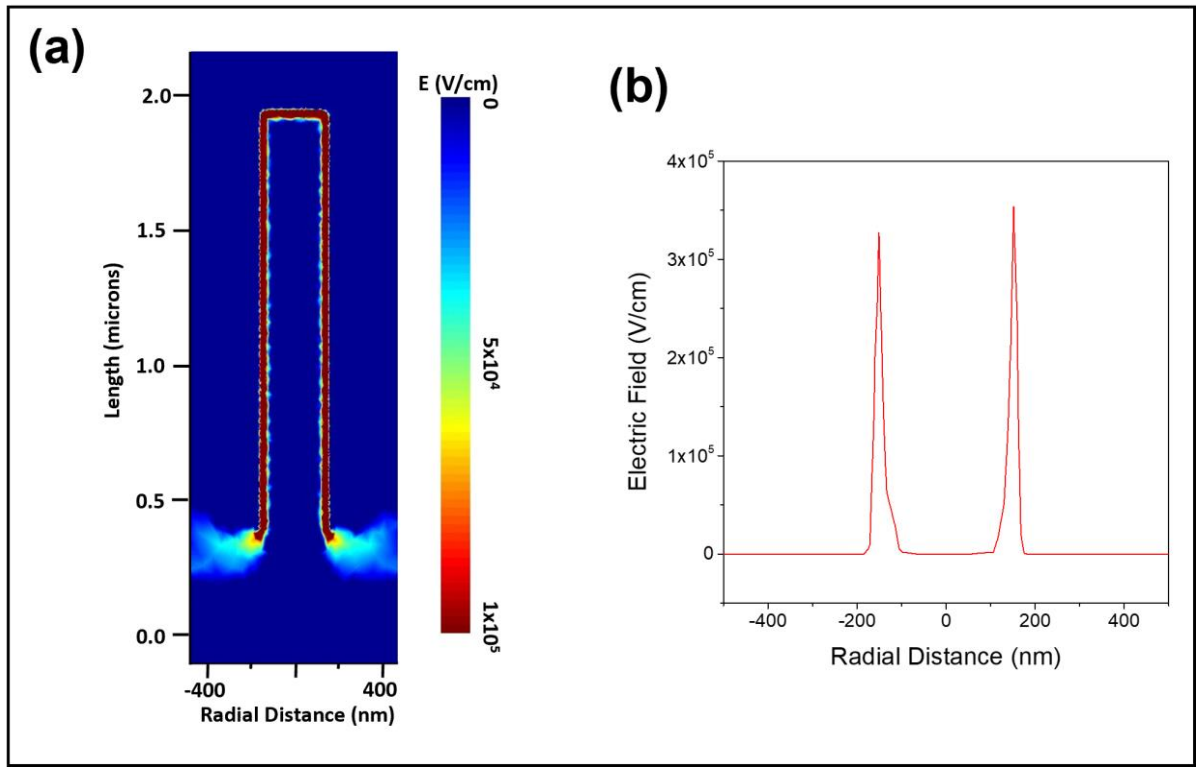

Fig 4. (a) Simulated built-in electric field across the radial p-n junction, and (b) shows that more than $3 \times 10^5$ V/cm of built-in electric field could be achieved across the radial junction at 0 V over very short distances. [Figure 4 has been reproduced (adapted) with permission from ref [9]. Copyright © 2021 John Wiley & Sons publications]

etched, InP nanowires which form the core, and ALD deposited AZO/ZnO, which form the shell. As shown in Figures 2(b) and 2(c), the deposition of an optimized thickness of AZO/ZnO can significantly enhance the absorption in bare InP nanowires due to reduced screening of incident electric field and an optical antenna effect. This enhancement allows nanowires with an oxide layer to achieve near-ideal $J_{sc}$ for a geometric filling ratio of less than 0.09 (see figure 2(b)).

Figure 2(d) shows simulated results on the effect of doping and surface recombination velocity for nanowires with a bulk carrier lifetime of 50 ps. Our simulation shows that even for a nanowire absorber lifetime of 50 ps, more than 23% efficiency can be achieved for a heavily doped p-InP nanowire core ($>3 \times 10^{18}$ cm$^{-3}$), if a moderately low interface recombination velocity of about $10^4$ cm/s could be achieved. Such high performance is a combined result of radial charge carrier separation boosted by a high built-in electric field across heavily doped p-InP and n-AZO/ZnO, and significantly reduced minority carrier recombination at the interface due to AZO/ZnO carrier selectivity and large valence band offset. A surface recombination velocity of $10^4$ cm/s for p-type InP nanowire is very realistic and can readily be achieved for InP nanowire through passivation, and therefore, the proposed solar cell design should achieve high efficiency in experiments, as shown below.

*B. Experimental Results*

Fabrication of radial junction nanowire solar cells started with optimization of large area electron beam lithography (see Figure 3(a)) and lift-off of chromium nanodisks, which acted as a mask for etching InP nanowires. The nanowires were formed through plasma etching. An SEM image of the etched nanowire array is shown in Figure 3(b). Figure 3(c) shows the highly conformal and uniform deposition of AZO/ZnO shell over the top-down etched nanowire array. The bulk lifetime of the substrate was about 88 ps (see Figure 3(d)), which reduced to below the resolution of the time resolve photoluminescence setup after nanowire etching due to surface damage, and the exact lifetime of the as-etched nanowire could not be measured. However, after plasma cleaning and deposition of the AZO/ZnO layer, a lifetime of about 53 ps could be measured.

Figure 3(e) shows that the measured efficiency of radial junction nanowire solar cells exceeded 16%, with active area efficiency of about 17% and a record-breaking active area short circuit current of 31.3 mA/cm$^2$. The high active area short circuit current is also reflected in measured external quantum efficiency shown in Figure 3(g), which exceeds more than 95% between the 550-750 nm regime. In addition, a planar device was fabricated using the same substrate to compare the effectiveness of radial junction nanowire solar cells. The efficiency of this device remained less than 1% even after several optimizations. A representative JV curve of the planar junction solar cell is shown in Figure 3(f).

We also performed a detailed loss analysis to understand the current limitation of our radial junction nanowire solar cells. It showed that by reducing metal shading, series resistance, and interface recombination, the proposed device could achieve more than 22% efficiency, which is similar to what was predicted by the simulation.

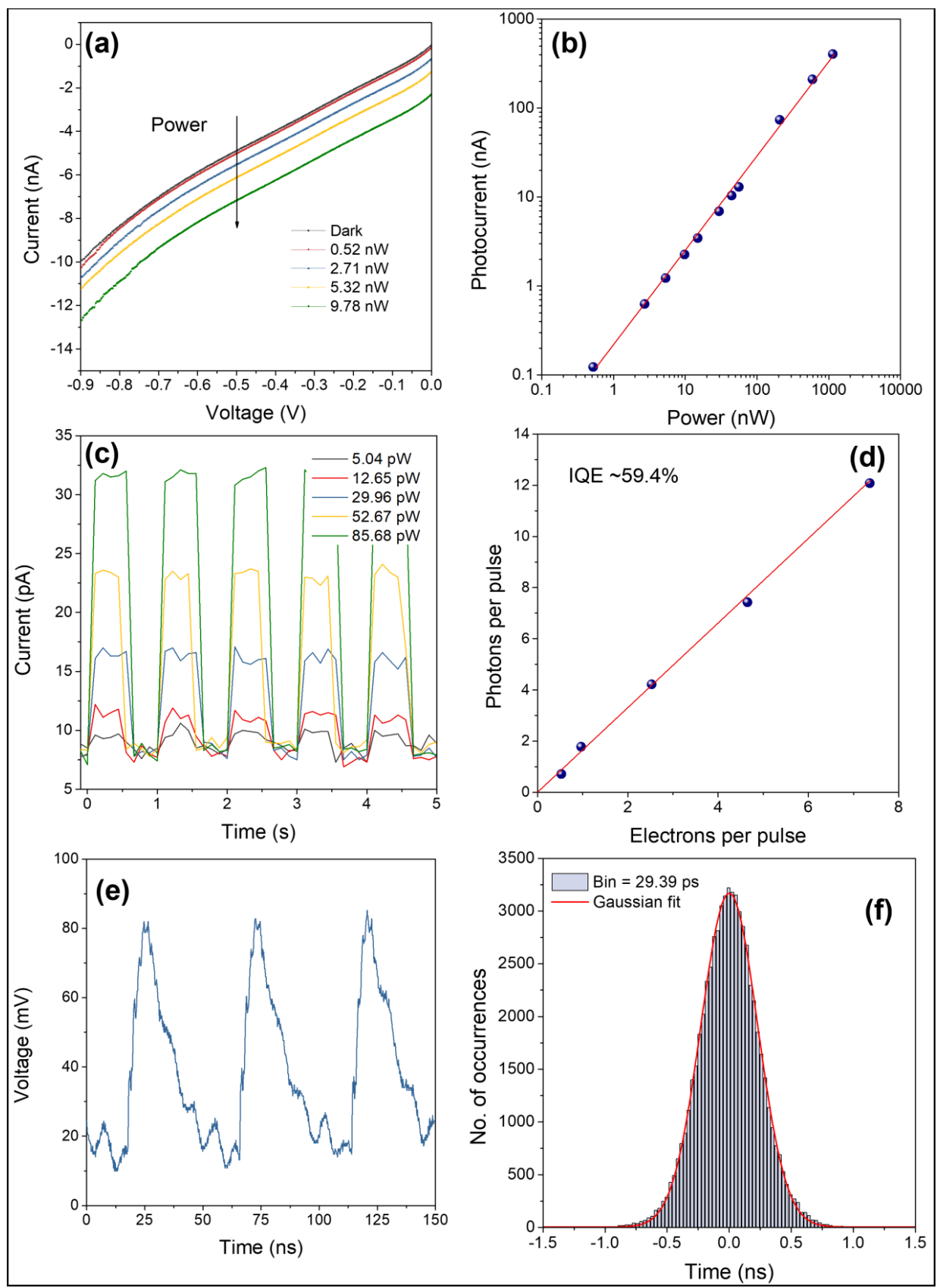

Fig 5. (a) Measured reverse IV curve of radial junction nanowire solar cell in the dark and under illumination. (b) Photocurrent vs illumination power plot showing almost linear response from a few pW to thousands of nW of illuminations. (c) Photocurrent measured at 300 K and 0 V bias for different ultra-low illumination powers. By attenuating 700 nm picosecond laser (6 ps pulse width) down to a single photon per pulse, the 5.04 pW is equivalent to the sub-single photon level. (d) The calculated internal quantum efficiency of a photodetector for a few photons per pulse illumination. (e) Output voltage pulse measured by oscilloscope using a 522 nm fs-laser excitation with a repetition rate of 20 MHz and power of 800 uW. (f) A Gaussian curve fitting of coincidence data measured at 80% of the pulse rising edge yields an estimated time jitter of 538 ps. [Figure 5 has been reproduced (adapted) with permission from ref [9]. Copyright © 2021 John Wiley & Sons publications]

## III. Nanowire Photodetector

The deposition of heavily doped, wide band gap, n-type AZO/ZnO over heavily doped p-InP leads to an extremely high built-in electric field across the radial junction. Also, the proposed structure could achieve near 100% absorption across a broad wavelength range in the visible regime. Therefore, by using a similar design for photodetection, we demonstrated extremely high sensitivity photodetection using these radial junction structures. In the next section, we summarize some of the results reported on the fabrication of a self-powered photodetector with sensitivity down to the single photon level.

Self-powered photodetection relies on generating a distinguishable signal difference between dark and the light regime at 0 V. Therefore, to achieve single photon detection, the dark current of the device should be sufficiently low that the signal generated by illumination of only a few picowatts of light could be differentiated from the dark current. Another requirement is that the device should be highly efficient in separating and collecting the electron-hole pair generated by an extremely small light signal. In the current case, we can perform a single photon level measurement because simultaneously a very small dark current and highly efficient charge carrier separation can be achieved in this architecture.

### A. Simulations

Figure 4 shows the simulated built-in electric field across the radial p-n junction formed by $p^+$-InP and $n^+$-ZnO/AZO. It is quite evident that the core-shell structure can achieve more than $3 \times 10^5$ V/cm built-in electric field within a very narrow depletion width of 40 nm, which allows extremely low dark current at 0 V, even when the p-n junction materials are of low quality (i.e., low bulk lifetime). Moreover, in the presence of light signal, electron-hole pairs generated within the p-n junction could be very efficiently separated to their respective collection electrodes, to contribute to photocurrent.

### B. Experimental Results

Figure 5(a) shows the photodetector performance in the dark and at different levels of illumination. An immediate conclusion that can be made from the figure is that the device has an extremely low leakage current under reverse bias conditions, which points to an excellent device behavior, and is a combined result of a high built-in electric field, low depletion region width, carrier selectivity, and passivation effect offered by radial heterojunction formed between AZO/ZnO and p-InP. Figure 5(b) shows that the photodetector has an extended dynamic range and could achieve photodetection down to a few picowatts with an almost linear increase in sensitivity.

To perform single photon measurement, the light signal was attenuated to a level with only one photon per pulse. For a 20 MHz pulsed laser and 700 nm wavelength, a single photon per pulse corresponded to 5.04 pW. Figure 5(c) shows that we could very reliably differentiate between the dark current and the current generated by the single photon pulse, with an excellent internal quantum efficiency of ~59.4% (see figure 5(d)).

Figure 5(e) shows the temporal response of the photodetector. Using a modified Gaussian fit, the 10 – 90 % rise time is estimated to be ~ 5.8 ns, and the FWHM is ~ 14.9 ns. This rapid response time points to high-speed photodetection even under zero-bias conditions. Another important temporal parameter of the photodetector is its timing jitter, which shows the uncertainty in the measurement of the photon arrival event due to fluctuations in the electrical signal. One of the primary causes of timing jitter is spatial variation, which causes a difference in photon collection timing depending on the distance from the collection electrode. A histogram representation of the number of occurrences of a specific type of waveform against time is plotted in Figure 5(f). The histogram was constructed to analyze the timing jitter by taking data at 80% of the pulse rising edge. The FWHM extracted from the Gaussian fit to the histogram is 538 ps. This relatively large timing jitter results from the relatively large area of our photodetector, which is 4 $mm^2$ in active area. This timing jitter could be significantly reduced by shrinking the size of the photodetector.

Finally, to show that our photodetector is of top quality we compare our photodetector with commercially available photodetectors in table 1. It is quite evident that our photodetector is comparable or on par with commercial photodetectors on almost all performance metrics. In particular, the noise-equivalent power of our photodetector is at least three orders of magnitude lower than those commercially available detectors. It is worth noting that our photodetector is originally fabricated from low quality absorber. With further improvement of the base absorber, it holds the potential to achieve even higher performance. This clearly shows that with future development, radial junction nanowire devices can provide a highly efficient and cost-effective alternative to established photodetection technologies for commercial applications.

## IV. Conclusion

In conclusion, we summarize some of our recent results on the design and fabrication of high-efficiency nanowire solar cells and self-powered single photon detectors by utilizing a radial junction geometry. We show both theoretical and experimental evidence that one can overcome the limitations of low-lifetime materials to achieve high-performance optoelectronics through this design. These results have huge implications for realizing low-cost photovoltaics and photodetectors using non-epitaxial growth techniques without compromising the overall device performance, which is currently not possible.

TABLE I. Comparison of proposed photodetector's performance metrices with commercially available photodetectors.

| Device Model | Vendor | Material | Responsivity (@700 nm) | Dark Current | Response Time | NEP |
|---|---|---|---|---|---|---|
| **This work** | N/A | InP | 0.354 A/W | 8.1 pA | 5.8 ns | $4.55 \times 10^{-17}$ |
| DET36A2 | Thorlabs | Si | 0.389 A/W | 350 pA | 14 ns | $1.6 \times 10^{-14}$ |
| DET10N2 | Thorlabs | InGaAs | 0.284 A/W | 1500 pA | 5 ns | $2 \times 10^{-14}$ |
| S5106 | Hamamatsu | Si | 0.5 A/W | 5000 pA | N/A | $1.6 \times 10^{-14}$ |
| G10899-003K | Hamamatsu | InGaAs | 0.26 A/W | 300 pA | N/A | $2 \times 10^{-14}$ |

## Acknowledgment

Australian Research Council and Australian National Fabrication Facility, ACT node are acknowledged for the financial support and access to facilities used in this work, respectively.